\documentstyle[psfig]{mn}
\newif\ifAMStwofonts
\ifoldfss
  \newcommand{\rmn}[1] {{\rm #1}}

  \ifCUPmtlplainloaded \else
    \NewTextAlphabet{textbfit} {cmbxti10} {}
    \NewTextAlphabet{textbfss} {cmssbx10} {}
    \NewMathAlphabet{mathbfit} {cmbxti10} {} % for math mode
    \NewMathAlphabet{mathbfss} {cmssbx10} {} %  "   "    "
  \fi
  \ifAMStwofonts
    \ifCUPmtlplainloaded \else
      \NewSymbolFont{upmath} {eurm10}
      \NewSymbolFont{AMSa} {msam10}
      \NewMathSymbol{\upi}     {0}{upmath}{19}
      \NewMathSymbol{\umu}     {0}{upmath}{16}
      \NewMathSymbol{\upartial}{0}{upmath}{40}
      \NewMathSymbol{\leqslant}{3}{AMSa}{36}
      \NewMathSymbol{\geqslant}{3}{AMSa}{3E}

    \fi
  \fi
\fi % End of OFSS

\ifnfssone
  \newmathalphabet{\mathit}
  \addtoversion{normal}{\mathit}{cmr}{m}{it}
  \addtoversion{bold}{\mathit}{cmr}{bx}{it}
  \newcommand{\rmn}[1] {\mathrm{#1}}

  \newmathalphabet{\mathbfit} % math mode version of \textbfit{..}
  \addtoversion{normal}{\mathbfit}{cmr}{bx}{it}
  \addtoversion{bold}{\mathbfit}{cmr}{bx}{it}
  \newmathalphabet{\mathbfss} % math mode version of \textbfss{..}
  \addtoversion{normal}{\mathbfss}{cmss}{bx}{n}
  \addtoversion{bold}{\mathbfss}{cmss}{bx}{n}
  \ifAMStwofonts
    \ifCUPmtlplainloaded \else
      \UseAMStwoboldmath
      \makeatletter
      \new@mathgroup\upmath@group
      \define@mathgroup\mv@normal\upmath@group{eur}{m}{n}
      \define@mathgroup\mv@bold\upmath@group{eur}{b}{n}
      \edef\UPM{\hexnumber\upmath@group}
      \new@mathgroup\amsa@group
      \define@mathgroup\mv@normal\amsa@group{msa}{m}{n}
      \define@mathgroup\mv@bold\amsa@group{msa}{m}{n}
      \edef\AMSa{\hexnumber\amsa@group}
      \makeatother
      \mathchardef\upi="0\UPM19
      \mathchardef\umu="0\UPM16
      \mathchardef\upartial="0\UPM40
      \mathchardef\leqslant="3\AMSa36
      \mathchardef\geqslant="3\AMSa3E
    \fi
  \fi
\fi % End of NFSS release 1

\ifnfsstwo
  \newcommand{\rmn}[1] {\mathrm{#1}}

  \DeclareMathAlphabet{\mathbfit}{OT1}{cmr}{bx}{it}
  \SetMathAlphabet\mathbfit{bold}{OT1}{cmr}{bx}{it}
  \DeclareMathAlphabet{\mathbfss}{OT1}{cmss}{bx}{n}
  \SetMathAlphabet\mathbfss{bold}{OT1}{cmss}{bx}{n}
  \ifAMStwofonts
    \ifCUPmtlplainloaded \else
      \DeclareSymbolFont{UPM}{U}{eur}{m}{n}
      \SetSymbolFont{UPM}{bold}{U}{eur}{b}{n}
      \DeclareSymbolFont{AMSa}{U}{msa}{m}{n}
      \DeclareMathSymbol{\upi}{0}{UPM}{"19}
      \DeclareMathSymbol{\umu}{0}{UPM}{"16}
      \DeclareMathSymbol{\upartial}{0}{UPM}{"40}
      \DeclareMathSymbol{\leqslant}{3}{AMSa}{"36}
      \DeclareMathSymbol{\geqslant}{3}{AMSa}{"3E}
    \fi
  \fi
\fi % End of NFSS release 2

\ifCUPmtlplainloaded \else
  \ifAMStwofonts \else % If no AMS fonts
    \def\upi{\pi}
    \def\umu{\mu}
    \def\upartial{\partial}
  \fi
\fi

\title{Late superhumps and the stream-disc impact in IY UMa}
\author[D. J. Rolfe et al.]
{Daniel J. Rolfe,$^1$ Carole A. Haswell,$^1$ and Joseph Patterson$^2$\\
  $^1$Department of Physics and Astronomy, The Open University, Walton
  Hall, Milton Keynes MK7 6AA\\
  $^2$Department of Astronomy, Columbia University, 538 W. 120th St.,
  New York, New York 10027} \date{Submitted July 2000}

\pagerange{\pageref{firstpage}--\pageref{lastpage}}
\pubyear{2000}

\begin{document}

\maketitle

\label{firstpage}

\begin{abstract}  
  We use the hot spot eclipse times of the newly discovered
  deeply-eclipsing dwarf nova IY UMa to trace out the shape of its
  disc during the late superhump era. We find an eccentric disc. We
  show that the brightness of the stream-disc impact region varies as
  expected with $|\Delta\vec{V}|^2$, where $\Delta\vec{V}$ is the
  relative velocity of the stream with respect to the velocity of the
  disc at the impact point.  We conclude that the hot spot is the
  source of late superhump light.
\end{abstract}

\begin{keywords}
  accretion discs -- binaries: close -- binaries: eclipsing - stars:
  individual: IY UMa -- stars: cataclysmic variables
\end{keywords}

\section{Introduction}
\label{intro}
IY UMa was identified as a dwarf nova type cataclysmic variable (CV)
in January 2000 (Uemura et al. 2000) when it went into superoutburst
showing strong superhumps and deep eclipses in its lightcurve.

Superhumps are luminosity variations with period, $P_{\rmn{sh}}$, a
few percent longer than the orbital period, thought to arise from the
interaction of the donor star orbit with a slowly progradely
precessing non-axisymmetric accretion disc. The eccentricity of the
disc arises because a 3:1 resonance occurs between the donor star
orbit and motion of matter in the outer disc. This can only occur in
systems with a sufficiently low mass ratio
($q=\frac{M_{donor}}{M_{wd}}$) that the 3:1 resonance radius is within
the tidal radius at which the disc is truncated by tidal forces
(Paczy\'{n}ski 1977).  The superhump period is then given by
$$\frac{1}{P_{\rmn{sh}}} = \frac{1}{P_{\rmn{orb}}} -
\frac{1}{P_{\rmn{prec}}}$$
where $P_{\rmn{prec}}$ is the disc
precession period and $P_{\rmn{orb}}$ is the orbital period.

A detailed study of the superoutburst of IY UMa was carried out by
Patterson et al. (2000) (hereafter P2000). IY UMa has orbital period
$P_{\rmn{orb}}=1.77$ hours. The system parameters were estimated in
P2000 as $M_{wd}=0.86\pm0.11 M_\odot$, $M_{donor}=0.12\pm0.02 M_\odot$ and
orbital inclination $i=86\fdg8 \pm 1\fdg5$.

\section{Observations}
\label{observations}
\begin{figure}
  \psfig{file=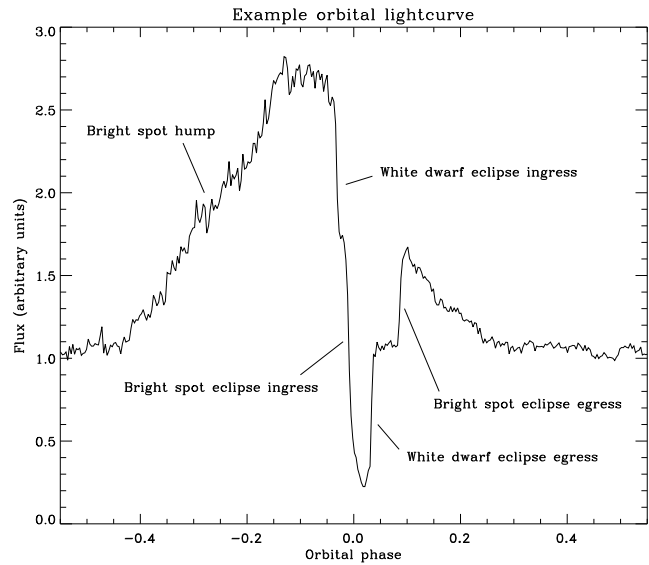,width=\columnwidth}
  \caption{  
    An example orbital lightcurve of IY UMa during the late superhump
    era which clearly shows the white dwarf and hot spot eclipse
    ingresses and egresses. This orbit has mid-eclipse at HJD
    2451578.69.
    }
  \label{orbcurve}
\end{figure}

Here we analyse V-light data covering the late superhump era, HJD
2451572--79, from the more extensive data presented in P2000. These
data comprise 22 eclipses with typical time resolution 20 to 30
seconds. Figure \ref{orbcurve} shows a representative orbital
lightcurve, where a strong orbital hump is visible as the stream disc
impact region (bright spot) comes into view before eclipse. The white
dwarf and hot spot eclipses are easily identified.

The entire dataset being analysed is shown in Figure
\ref{lightcurves}a, and is discussed in Section \ref{discussion}.

\section{Tracing the eccentric disc shape}
\subsection{Shadow method}
\label{shadowmethod}
\begin{figure}
  \psfig{file=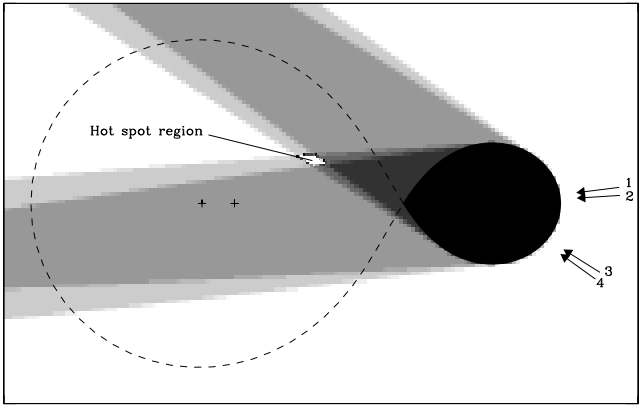,width=\columnwidth}
  \caption{  
    ``Shadows'' cast by the donor star at the beginning and end of hot
    spot ingress (observer looking from 1 and 2 respectively) and at
    start and end of hot spot egress (3 and 4) constrain the location
    of the hot spot in the orbital plane to a diamond shape.  }
  \label{shadows}
\end{figure}

The eclipse of the prominent hot spot (Figure \ref{orbcurve}) provides
an opportunity to probe the disc shape. Assuming the surface of the
donor star is described by its Roche potential surface, and if the
mass ratio, $q$, and orbital inclination, $i$, are known, the region
of the orbital plane eclipsed by the donor at any particular orbital
phase can be calculated. Figure \ref{shadows} shows how these
``shadows'' cast by the donor star at phases corresponding to the
start and end of hot spot eclipse ingress and egress constrain the
location of the hot spot in the orbital plane to a diamond shaped
region. This method has been used before, e.g. Wood et al.  (1986).
The hot spot is the point where the accretion stream impacts the edge
of the disc, so determining the hot spot location determines the
location of the disc edge.  Assuming the disc precesses at the beat
period between the orbital and superhump periods, the geometry shown
in Figure \ref{geometry} was used to trace out the outline of the disc
edge using those eclipses where both ingress and egress of hot spot
eclipse could be measured.

\subsection{Stream trajectory method}
\label{streammethod}
\begin{figure}
  \psfig{file=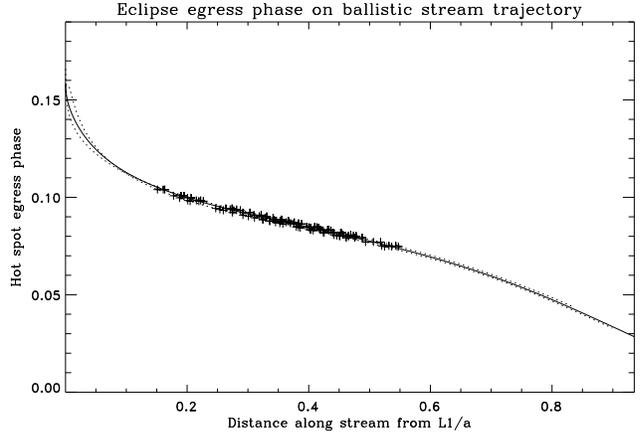,width=\columnwidth}
  \caption{  
    Orbital phase of eclipse egress for points on a ballistic stream
    trajectory calculated assuming the shape of the donor star to be
    described by its Roche potential surface. The solid curves
    represent a stream started from the L1 point with
    $(V_x,V_y)$=$(-C_s,0)$ (see Section \ref{streammethod}) while the
    dotted curves are streams started with $(V_x,V_y)$=$(-C_s,\pm
    C_s)$. The crosses are the measured egress phases at their
    inferred positions along the stream.  }
  \label{streamcontact}
\end{figure}

\begin{figure}
  \psfig{file=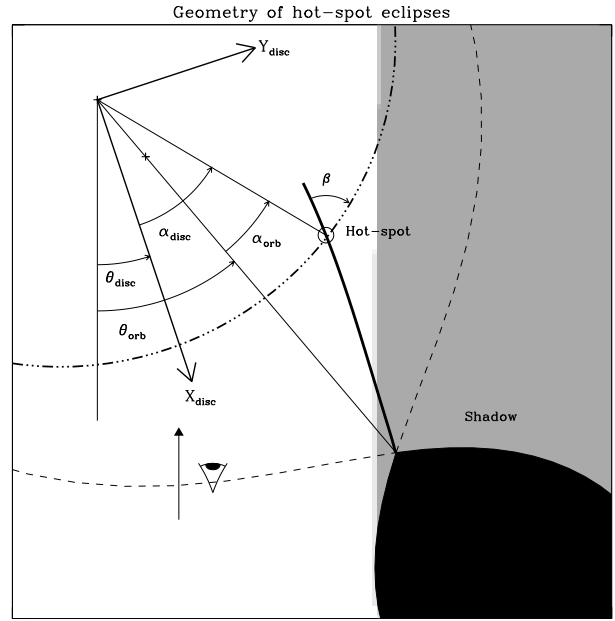,width=\columnwidth}
  \caption{  
    The geometry of the system used to study the hot spot eclipses.
    This figure shows the situation at orbital phase 0.11, shortly
    after a hot spot eclipse. The disc coordinates
    $(X_{disc},Y_{disc})$ are centred on the white dwarf. The stream
    impacts the disc edge (thick dot-dashed outline) at the hot spot,
    at angle $\alpha_{orb}$ to the line of centres of the two stars,
    which itself makes an angle $\theta_{orb}$ with the line of sight.
    The X-axis of the disc makes an angle $\theta_{disc}$ with the
    line of sight, and so the position of the stream-disc impact in
    the precessing disc frame makes an angle
    $\alpha_{disc}=\alpha_{orb}+\theta_{orb}-\theta_{disc}$ with the
    disc X-axis. $\beta$ is approximately the angle between the stream
    and disc velocities at the point of impact.}
  \label{geometry}
\end{figure}

For some eclipses, only the egress of hot spot eclipse could be
measured, and so the technique described above could not be employed
for those eclipses. However there is an alternative technique for
determining the location of the hot spot which can be employed in this
situation (Hessman et al. 1992). Since the hot spot lies on the
accretion stream, and the times of hot spot egress depend on the
distance of the hot spot from the line of centres of the two stars, a
ballistic trajectory for the accretion stream can be used to infer the
position of the hot spot along the stream. Hence the location of the
hot spot and the disc edge can be determined. Figure
\ref{streamcontact} shows the variation of eclipse egress phase along
a ballistic stream trajectory.

The uncertainty in the time of each point of contact translates mostly
to an error in the radial distance of the hot spot from the white
dwarf for an assumed $q$. Azimuthal error was estimated by considering
three slightly different stream trajectories with initial velocities
at the inner Lagrange point (with axes as shown in Figure
\ref{discshape}) of $(V_x,V_y)$=$(-C_s,0)$, $(-C_s,C_s)$ and
$(-C_s,-C_s)$. $C_s$ is the sound speed at the L1 point estimated
using $$C_s\approx10^4 m s^{-1} \sqrt{\frac{T}{10^4K}}$$
where the
photospheric temperature $T$ comes from the stellar models by Baraffe
(Baraffe et al. 1998). The different hot spot positions coming from
these three slightly different trajectories provide the small
near-azimuthal error bars on the egress and ingress locations in
Figure \ref{discshape}.

\subsection{Combining the methods}

The disc radius determined using the stream trajectory method depends
on whether the start, middle or end of egress is used. We find that
for those eclipses where both ingress and egress were measured, using
the mid-egress times to locate the hot spot on the accretion stream
gives results consistent with the robust ``shadow'' method whose
results completely constrain the location of the hot spot in the
orbital plane. We therefore use the shadow method to determine hot
spot location and disc radius for those eclipses where both hot spot
ingress and egress are measurable, and the stream trajectory method
with mid-egress time for the remaining eclipses.

We use the white dwarf mid-eclipse ephemeris
$T_{mid} = HJD~2451570.85376(2)~+~0.07390906(7)E$ and late superhump
maximum ephemeris $T_{mid} = HJD~2451571.731(3)~+~0.07558(6)E$ from
P2000 to calculate orbital phase and superhump phase. The beat phase
between orbital and superhump phase, interpreted as disc precession
phase $\phi_{prec}$, is defined as zero when superhump maximum and
mid-eclipse coincide.  The angle $\theta_{disc}$ (see Figure
\ref{geometry}) is then $2\pi\phi_{prec}$.

\section{Results}
\label{results}

\begin{figure}
  \psfig{file=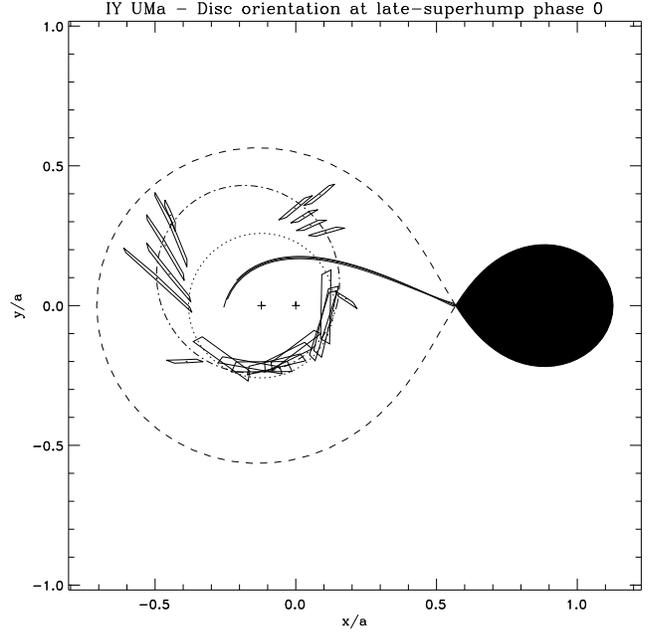,width=\columnwidth}
  \caption{  
    The disc shape traced out from the hot spot ingress and egress
    times. The disc orientation shown is that which would occur at
    superhump maximum. Thin regions with large radial extent were
    determined using the stream trajectory method (Section
    \ref{streammethod}). The other regions, which are azimuthally
    extended (due to the azimuthal extent of the hot spot region and
    the relative motion of hot spot and disc frame during hot spot
    eclipse) were determined using the ``shadow'' method (Section
    \ref{shadowmethod}). The dot-dashed outline is the fitted disc
    shape and the dotted circle has radius $r_{circ}$ (the
    circularization radius for mass ratio $q=0.14$) and is plotted to
    make the non-axisymmetry of the disc easier to see.  }
  \label{discshape}
\end{figure}

The average disc radius $R_{disc}$ in units of orbital separation $a$
is $<R_{disc}/a>=0.32$ with a spread of 0.08. The disc shape is shown
in Figure \ref{discshape} and is clearly not centred on the white
dwarf. The shape is similar to that found in OY Car by Hessman et al.
(1992).
 
\begin{figure}
  \psfig{file=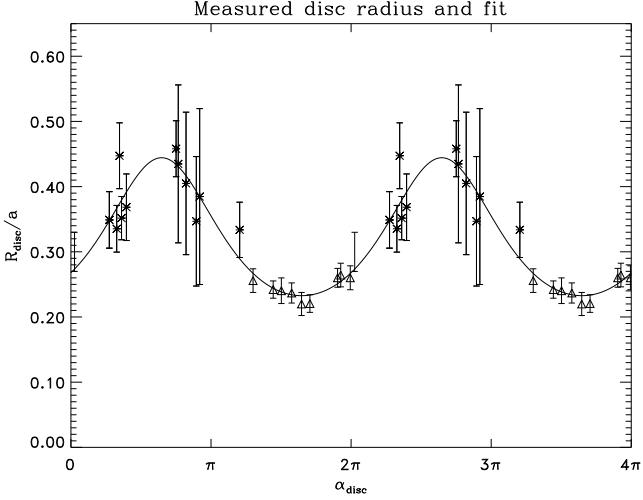,width=\columnwidth}
  \caption{    
    The disc radius as a function of $\alpha_{disc}$ (see definition
    in Figure \ref{geometry}). Points marked with triangles were
    determined using the ``shadow'' method (Section
    \ref{shadowmethod}), while points marked with stars were
    determined using the stream trajectory method (Section
    \ref{streammethod}).}
  \label{discfit}
\end{figure}

A good fit to the disc shape was achieved using an ellipse with one
focus at the white dwarf, semi-major axis $a_{disc}$ and eccentricity
$e$, described by
$$R_{disc}=\frac{a_{disc}(1-e^2)}{1-e\,cos(\alpha_{disc}-\alpha_0)}.$$
$\alpha_0$ is the angle in the precessing disc frame corresponding to
maximum disc radius. The resulting fit has $a_{disc}=0.34a \pm 0.06a$,
$e=0.31 \pm 0.17$ and $\alpha_0=117 \pm 33^\circ$ and is shown in
Figure \ref{discfit}.  The errors in our timing determinations and
hence deduced geometry (particularly when using the stream trajectory
method) are rather pessimistic; our determinations of $a_{disc}$, $e$
and $\alpha_0$ are robust to adopting instead more optimistic error
estimates.

The disc is within the tidal cut-off radius $r \sim 0.9R_L=0.50a$.
The largest radius part reaches the location of the 3:1 resonance at
$r \sim 0.46a$ (Osaki 1996), providing a possible explanation for the
maintenance of disc eccentricity at this stage after the
superoutburst. The smallest radius coincides with the circularization
radius.

\section{Discussion}
\label{discussion}

The disc shape we find is similar to that found for OY Car by Hessman
et al. (1992), except that we find the minimum radius at disc azimuth
$\alpha_{disc}=-63^\circ$ while in OY Car it is around
$\alpha_{disc}=0$. This means the smallest radius of the disc is lined
up with the donor star at superhump phase -0.18 in IY UMa while this
occurs close to superhump phase 0 in OY Car. This has implications for
the viability of the hot spot as the source of the superhump light
during the late superhump era. Superhumps caused by the variation in
hot spot brightness as the stream impacts the disc at varying depth in
the white dwarf potential well were first suggested by Vogt (1981).
This theory has generally been rejected as an explanation for the
common superhumps occurring during the decline from superoutburst
maximum in SU UMa stars, but the possible link between this model and
\emph{late} superhumps (which appear during very late decline of the
superoutburst) has been suggested several times, e.g. by Whitehurst
(1988) and by Rolfe, Haswell \& Patterson (2000) (hereafter RHP) after
a study of persistent superhumps in the novalike V348 Pup. See RHP for
more discussion.

\begin{figure}
  \psfig{file=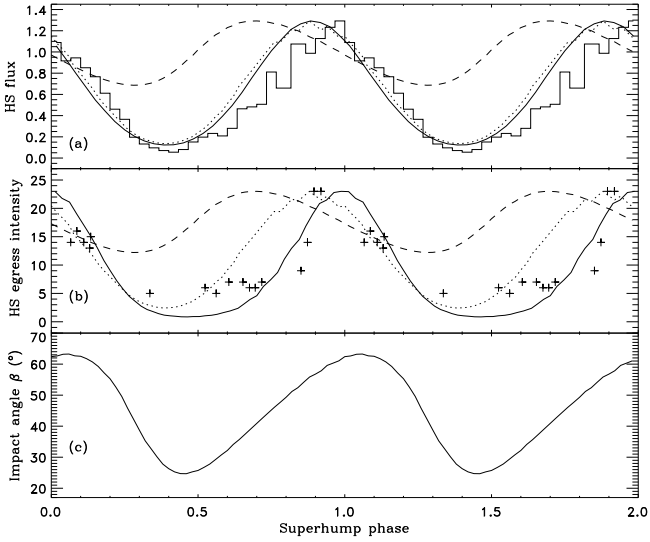,width=\columnwidth}
  \caption{
    (a) The histogram shows the peak bright spot modulation, obtained
    by subtracting the normalized average orbital curve from the
    normalized data and then extracting the data for the orbital phase
    range 0.85--0.95 and folding on superhump phase. (b) shows the
    measured hot spot eclipse egress intensity $\Delta I_{hse}$
    (crosses).  The dashed curve in (a) and (b) is the kinetic energy
    of the stream at the hot spot, while the dotted curve is the
    kinetic energy of relative motion of the stream and disc at the
    hot spot.  The solid curves are the predicted hot spot peak and
    egress intensities combining our simple 3D model for the hot spot
    structure with the relative KE model for hot spot brightness. See
    Section \ref{discussion}. The latter three curves were predicted
    from the fitted disc shape and scaled in flux to have the same
    maximum as the measured data. (c) shows the angle $\beta$ between
    the stream and disc velocities at the impact point.}
  \label{hotspotflux}
\end{figure}
  
In Figure \ref{hotspotflux}a we show a measure of the orbital hump
peak modulation while in Figure \ref{hotspotflux}b we show the hot
spot flux uncovered on egress of the hot spot eclipse ($\Delta
I_{hse}$ from P2000). The variation in hot spot brightness,
$F_{hotspot}$, predicted by our fitted disc shape is shown in Figures
\ref{hotspotflux}a \& b, calculated using three models described
below.

\begin{figure}
  \psfig{file=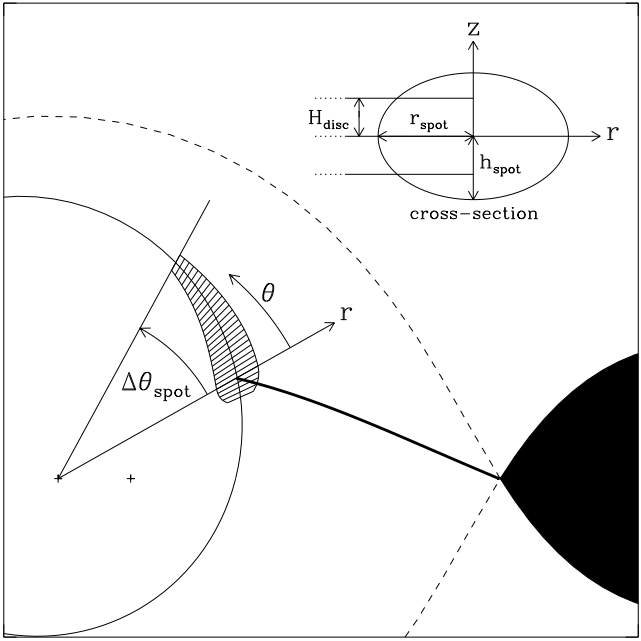,width=\columnwidth}
  \caption{
    The assumed 3D geometry of the hot spot seen looking down onto the
    orbital plane in the $-z$ direction. The hot spot is split up into
    ``rings'' in the $rz$-plane which are ellipses centred on the
    elliptical disc edge with axis in the $r$ direction of $r_{spot}$
    and vertical axis $h_{spot}$. Both $r_{spot}$, $h_{spot}$ and the
    hot spot brightness decrease as
    $e^{-\theta^2/\Delta\theta_{spot}^2}$ as the stream merges with
    the disc, and the stream impact is rounded off with a hemisphere
    for $\theta<0$.}
  \label{brightspotstructure}
\end{figure}

\begin{figure*}
  \begin{minipage}[t]{\columnwidth}
    \psfig{file=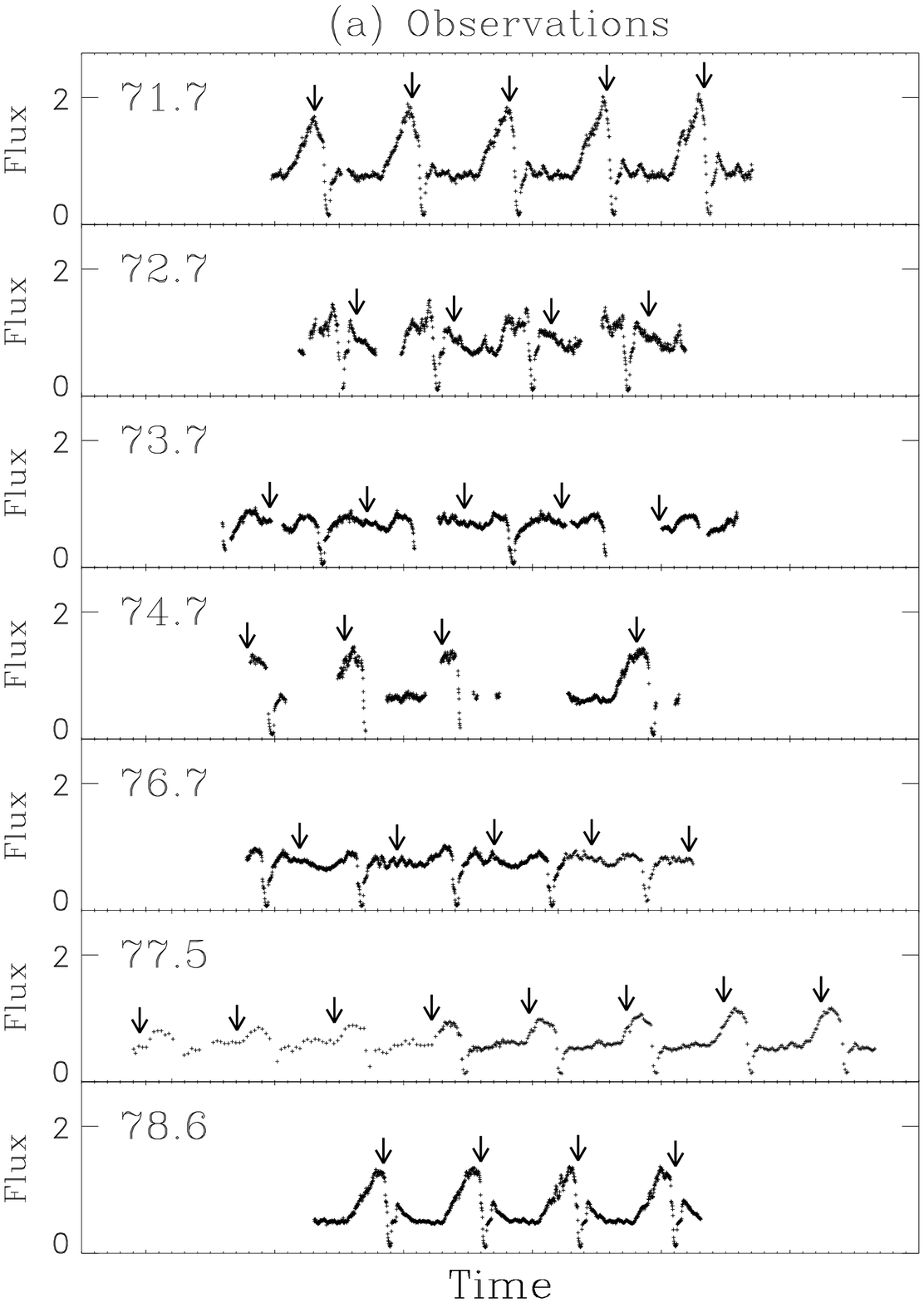,width=\columnwidth}
  \end{minipage}
  \hspace{4mm}
  \begin{minipage}[t]{\columnwidth}
    \psfig{file=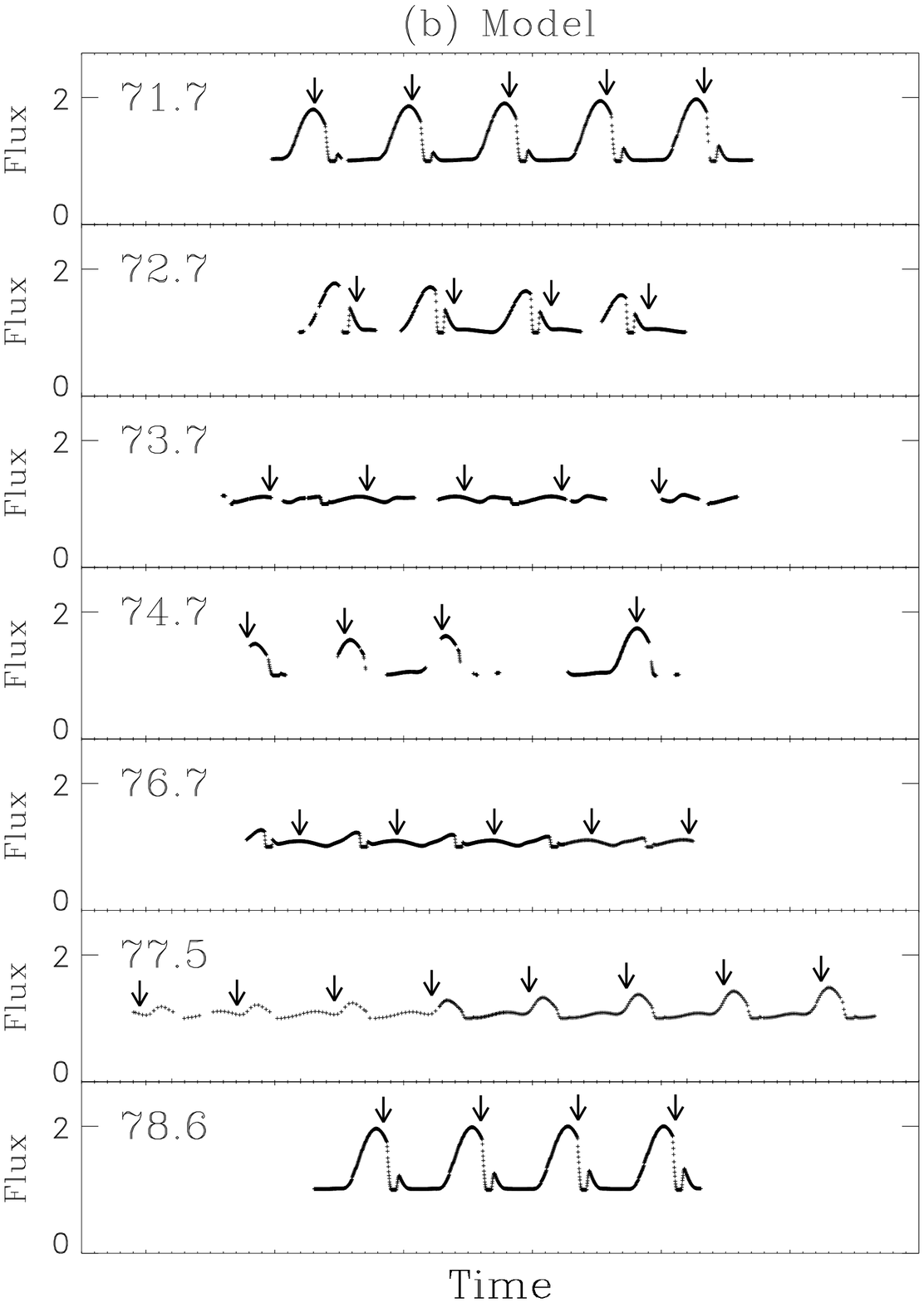,width=\columnwidth}
  \end{minipage}

  \vspace*{3mm}
  \begin{minipage}[t]{12cm}
    \psfig{file=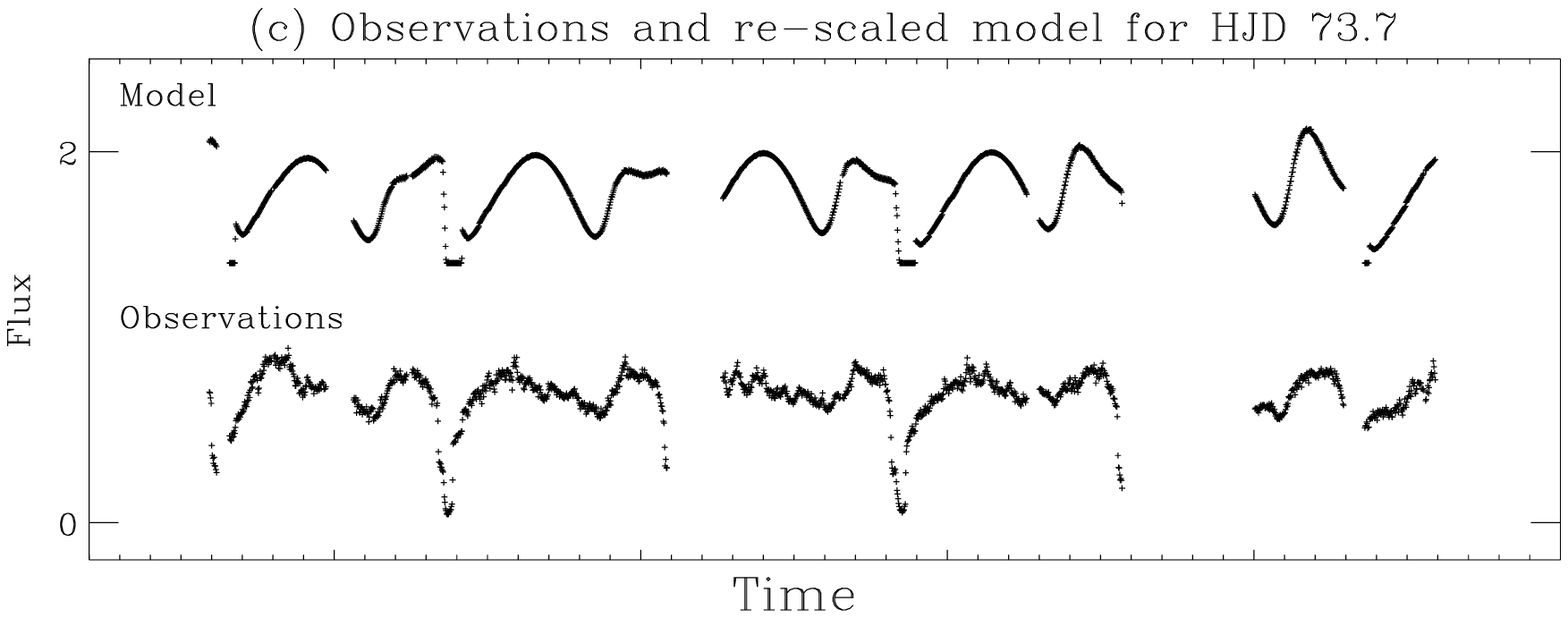,width=12cm}
  \end{minipage}
  \caption{
    (a) shows all 7 lightcurves from the late superhump dataset
    analysed in this paper. The date of the start of each lightcurve
    is shown in the top left of each plot as HJD-2451500. Arrows
    indicate times of late superhump maximum according to the
    ephemeris in P2000. Note the significant variation in the
    amplitude of the orbital hump over the 7 days. (b) Shows our model
    of the bright spot lightcurve for the same time series as (a). The
    model assumes the intrinsic hot spot brightness to vary as the
    kinetic energy of relative motion of stream and disc at impact,
    while a simple model for the 3D structure of the hot spot enables
    us to predict its visibility as a function of orbital phase and
    precession phase.  See Section \ref{discussion}. (c) shows the
    model data from from the 3rd panel of (b) scaled in flux by a
    factor of 5 and shifted up by 1.4 flux units, along with the
    corresponding observed data. This shows how the shape of the model
    lightcurve is morphologically close to that of the observations,
    although the intrinsic hot spot brightness predicted by our model
    is too low at the disc precession phase. Note that our model does
    not include the deep white dwarf eclipse seen in the
    observations.}
  \label{lightcurves}
\end{figure*}

\begin{enumerate}
\item The dashed curve in Figure \ref{hotspotflux}a \& b is calculated
  using $F_{hotspot}\propto 1/r$ where $r$ is the radius of the disc
  at the hot spot. This assumes the energy released at the hot spot
  varies as its depth in the white dwarf gravitational potential.
\item The maximum energy which could be released at impact is the
  kinetic energy of the relative motion of the stream and disc at the
  hot spot. If the stream velocity vector at the point of impact is
  $\vec{V}_{stream}$ and the disc velocity at this point is
  $\vec{V}_{disc}$ then the kinetic energy per unit mass of the
  relative velocity is
  $KE=\frac{1}{2}(\vec{V}_{disc}-\vec{V}_{stream}).(\vec{V}_{disc}-\vec{V}_{stream})$.
  We approximate
  $$|\vec{V}_{disc}|=\sqrt{GM_{wd}\left(\frac{2}{r}-\frac{1}{a_{disc}}\right)}$$
  at the impact point with direction parallel to the disc edge in the
  white dwarf frame. This is the correct velocity for an elliptical
  orbit around a point mass.  This model is plotted as the dotted
  curve in Figures \ref{hotspotflux}a \& b.
\item The solid curves in Figures \ref{hotspotflux}a \& b were
  produced by assuming the total hot spot flux behaves as described in
  the previous model, and considering a simple model for the 3
  dimensional structure of the hot spot, shown in Figure
  \ref{brightspotstructure}. The hot spot has an elliptical cross
  section in the $r-z$ plane with axis of size $r_{spot}$ in the
  radial direction and $h_{spot}$ in the vertical direction (see inset
  in Figure \ref{brightspotstructure}). $r_{spot}$ and $h_{spot}$ and
  the spot brightness decrease downstream from the initial impact as
  $e^{-\theta^2/\Delta\theta_{spot}^2}$. Upstream of the impact the
  hot spot surface is rounded off with a hemisphere of uniform
  brightness equal to that at the initial impact. The angular extent
  of the hot spot region is set by $\Delta\theta_{spot}=\Delta\theta_0
  a_{disc}/r_0$, where $r_0$ is the disc radius at the impact point.
  This keeps the arc length of the hot spot region roughly constant as
  the eccentric disc precesses. The disc thickness is
  $2H_{disc}=0.02a$ and any region of the hot spot surface within the
  disc is considered to be obscured completely. The total flux coming
  from an area element on the hot spot surface is foreshortened
  according to its area projected in the direction of the observer.
  This simple parametrized model for the hot spot structure enables us
  to model the full hot spot lightcurve, taking into account the
  visibility of the hot spot at any orbital phase and disc precession
  phase. The model lightcurve is shown in Figure \ref{lightcurves}b.
  We use $\Delta\theta_0=36^\circ$ which is the average azimuthal
  extent of the hot spot regions determined from the eclipse timings
  as described in Section \ref{shadowmethod}. Noting that we see no
  eclipse of the white dwarf by the hot spot region, we determine the
  maximum possible value of $h_{spot}$ for $i=86\fdg8 \pm 1\fdg5$
  to be $0.013a \pm 0.006a$. Therefore we use
  $r_{spot}(\theta=0)=h_{spot}(\theta=0)=0.013a$ at the point of
  impact.

\end{enumerate}

The phasing of the hot spot brightness predicted by model (i) does not
agree with either of the two measures of hot spot variation in Figures
\ref{hotspotflux}a \& b, with the peak in predicted brightness being
about 0.25 earlier than the measured peak. The fractional amplitude in
the predicted variation is about half that of the measured variation:
in Figures \ref{hotspotflux}a \& b the predicted curve has simply been
scaled to have the same maximum value as the observed hot spot curves.

In Figure \ref{hotspotflux}c we also show the variation of the angle
$\beta$ between the stream and disc velocities at the point of impact
(approximately equal to $\beta$ in Figure \ref{geometry}). The hot
spot emission should arise primarily from dissipation of kinetic
energy as the infalling stream impacts the disc edge and merges with
the Keplerian motion of the disc. The available kinetic energy
increases as the angle $\beta$ between the velocity vectors of the
stream and disc flows increases.  Figure \ref{hotspotflux}c clearly
shows $\beta$ to be greatest around the observed maximum of the hot
spot intensity, which coincides with superhump maximum. This effect is
taken into account in model (ii) where the maximum kinetic energy
available at the impact point is calculated and depends on the
directions and magnitudes of the stream and disc velocities. The
shape, phasing and fractional amplitude of the curve for model (ii)
shows much better agreement with the hot spot brightness measured in
Figure 7a. The predicted curve also agrees well in phase and
fractional amplitude with the measure of hot spot flux in Figure
\ref{hotspotflux}b, although the hot spot flux between superhump phase
$\sim$0.6 and 0.9 is much lower than predicted.

Our final model, (iii), which combines the relative kinetic energy
model with a treatment of the hot spot visibility, does not differ
from the previous model in its prediction for the peak flux of the
orbital hump (Figure \ref{hotspotflux}a). This is to be expected since
at the peak of the orbital hump we should be seeing the whole length
of the bright spot clearly, so the visibility at this orbital phase
should not be very sensitive to small changes in orientation through
the disc precession cycle. The hot spot flux measured using the hot
spot eclipse egress is different from the predicted variation in
intrinsic hot spot flux.  This is because at the orbital phase of hot
spot egress, we are seeing the impact region roughly end-on, and at
this orientation the fraction of the hot spot flux reaching the
observer is sensitive to the exact orientation of the hot spot. The
result is that the predicted variation (solid curve in Figure
\ref{hotspotflux}b) rises to maximum later than the intrinsic
variation, but decreases at about the same time as the intrinsic
variation.  This produces a more flat-bottomed curve which agrees
better in shape with the observed egress flux than models (i) and
(ii). The flux around egress flux minimum is too low in our models,
but this is not surprising; our model for the structure of the impact
region is a very simple one. It is also possible that the hot spot
emission is strongly anisotropic due to the complicated structure of
shocks and contact discontinuities we expect to be present at the
impact.

Figure \ref{lightcurves}a shows the evolution of the IY~UMa lightcurve
throughout the late superhump era. The important features to note are
how strongly the amplitude of the orbital hump varies, and how the
late superhump is very weak, in many cases undetectable, when it does
not occur near the orbital hump. Strong variation in hump amplitude
was also seen towards the end of a more recent superoutburst of IY~UMa
(Uemura 2000). The combined lightcurve of the late superhump and the
orbital hump is not simply a sum of the two, it is some non-linear
combination, with the combined hump being extremely strong when
orbital hump and late superhump coincide, while both orbital hump and
late superhump are weak when they are not coincident.  This is exactly
what we expect if the stream-disc impact is the source of the late
superhump. The hot spot is modulated at the late superhump period, as
the impact geometry of the stream and disc varies.  Consequently, the
orbital hump, which is just the emission from the hot spot coming into
view, is modulated on the disc precession period, being strongest at
late superhump maximum. The late superhump is very weak away from the
orbital hump since away from the orbital hump the bright spot is on
the far side of the disc and hence has a low visibility. Our model hot
spot lightcurves produced using model (iii) are shown in Figure
\ref{lightcurves}b. Ignoring the stochastic flickering in the
observations (which is strongest in the second panel, HJD 72.7), and
the deep white dwarf eclipse in the observations not included in the
model, our simple model does a convincing job of reproducing the
observed bright spot behaviour, although the peak of the orbital hump
when it occurs close to late superhump maximum is sharper in the
observations than in our model.

The intrinsic bright spot flux in the lightcurve in the third panel
(HJD 73.7) appears to be lower than the observed flux, which make is
difficult to see how the shape of this curve agrees with observations.
Figure \ref{lightcurves}c shows the model and observations for this
curve, with the model scaled by a factor of 5. The good agreement
between the shape of the model and the observations is clear,
particularly the double humped nature of each orbit curve and the
flattish top to the orbital hump before some of the eclipses. More
detailed modeling is necessary to explain the low intrinsic brightness
of the hot spot in this region of the disc precession cycle. The
relative height of the hot spot peak above the disc needed a bit of
fine-tuning to achieve this match between the model and observed
lightcurves, and $r_{spot}(\theta=0)$ had to be less than about twice
$h_{spot}(\theta=0)$.

It is difficult to avoid the conclusion that the late-superhump light
in IY UMa is predominantly coming from the hot spot, whose brightness
is modulated at the superhump period as a result of the varying impact
geometry of the stream and the disc. The disc shape and orientation
found in OY Car (Hessman et al. 1992) suggests that the late
superhumps in that system may also originate from the hot spot. An
analysis of the OY Car observations like that in our Figure
\ref{hotspotflux} would be valuable. Murray (1996) performed SPH
simulations which support the hot spot as the source of late superhump
light, but radiative processes were not modelled.  More detailed
hydrodynamic simulations concentrating on the stream-disc impact
region have been carried out (e.g. Armitage \& Livio 1998), but there
has been no study of the hot spot in such detail for non-circular
discs. Such work would be valuable in the light of these observations.

We haved assumed the shape of the disc to be fixed and constant in the
precessing disc frame, while our observations of V348 Pup (RHP) and
SPH simulations (Haswell, King, Murray \& Charles 2000) suggest the
disc shape and size changes during the superhump cycle. The disc shape
we find represents the disc radius seen by the accretion stream as a
function of superhump phase, and any effect of changing shape should
have little effect on the angle $\beta$ at which the stream impacts
the disc since the shape is traced out between radius measurements
which are closely spaced in superhump phase.

The transition from the common superhumps which appear during the
decline from maximum of the superoutburst to the late superhumps
during the very late decline of the superoutburst is easy to explain
using a minor modification to the tidal-thermal-instability model of
Osaki (1989) proposed by Hellier (2000). Hellier suggests that the
tidal and thermal instabilities are uncoupled: the disc can drop out
of the thermal high state at the end of the superoutburst while
remaining tidally eccentric. At this point, the disc viscosity
dramatically reduces and so the viscous dissipation due to tidal
stressing which causes normal superhumps will be similarly reduced,
allowing the late superhumps which were drowned out during the
superoutburst to be seen.

\section{Summary}
\label{summary}

The timing of hot spot eclipses during the late superhump era reveals
the disc shape, which proves to be non-axisymmetric.

The inferred shape and orientation of the disc and a simple treatment
of the kinetic energy available at the stream-disc impact enable us to
calculate the hot spot brightness throughout the superhump cycle. The
resulting variation in predicted hot spot brightness matches the
observed variation well, and is in very close agreement with the
superhump modulation. Combining these results with a simple model for
the 3D structure of the hot spot enables our model to reproduce the
observed lightcurves during the late superhump era.

We conclude that the stream-disc impact region is the source of
late superhump light in IY UMa.

\section{Acknowledgements}

The authors are very grateful to Jonathan Kemp, Tonny Vanmunster and
Bob Fried who contributed the data analysed here. We acknowledge the
data analysis facilities at the Open University provided by the OU
research committee and the OU computer support provided by Chris
Wigglesworth.  We thank Andrew King and Ulrich Kolb for helpful advice
and discussion, and Rick Hessman whose suggestion prompted this work.
The comments from the anonymous referee are also appreciated. DJR is
supported by a PPARC studentship.  CAH gratefully acknowledges support
from the Leverhulme Trust F/00-180/A.

\label{lastpage}

\end{document}